\documentclass[prc,twocolumn,showpacs,preprintnumbers,amsmath,amssymb]{revtex4-1}
\usepackage{graphicx}
\usepackage{bm}
\begin{document}
\title{Truncation scheme of time-dependent density-matrix approach}
\author{Mitsuru Tohyama} 
\affiliation{
Kyorin University School of Medicine, Mitaka, Tokyo
181-8611, Japan}
\author{Peter Schuck}
\affiliation{Institut de Physique Nucl$\acute{e}$aire, IN2P3-CNRS,
Universit$\acute{e}$ Paris-Sud, F-91406 Orsay Cedex, France}
\affiliation{Laboratoire de Physique et de Mod\'elisation des Milieux Condens\'es, CNRS 
 et, Universit\'e Joseph Fourier, 25 Av. des Martyrs, BP 166, F-38042 
 Grenoble Cedex 9, France}
\begin{abstract}
A truncation scheme of the Bogoliubov-Born-Green-Kirkwood-Yvon hierarchy
for reduced density matrices, where a three-body density matrix is 
approximated by the antisymmetrized products of two-body density matrices, is proposed.
This truncation scheme is tested for three model hamiltonians. It is shown that
the obtained results are in good agreement with the exact solutions.
\end{abstract}
\maketitle
\section{Introduction}
The equations of motion for reduced density matrices have a coupling scheme known as 
the Bogoliubov-Born-Green-Kirkwood-Yvon (BBGKY) hierarchy where an $n$-body density matrix couples to
$n$-body and $n+1$-body density matrices. An approximation of a two-body density matrix with the antisymmetrized products of
one-body density matrices in the equation of motion for the one-body density matrix gives the time-dependent Hartree-Fock theory (TDHF).
If we approximate a three-body density matrix with the antisymmetrized products of one-body and two-body density matrices in the
equation of motion for the two-body density matrix, we can truncate the BBGKY hierarchy and obtain the coupled equations of motion
for the one-body and two-body density matrices. This is the time-dependent density-matrix theory (TDDM) \cite{WC,GT}.
A stationary solution of TDDM gives the correlated ground state and its small amplitude limit (STDDM) is an extension of the random-phase
approximation (RPA) where the effects of ground-state correlations and the coupling to two-body amplitudes are included.
TDDM and STDDM have been applied to model hamiltonians \cite{TTS,toh07} and realistic ones \cite{toh07,pfitz,lacroix,toh12}.
In some cases TDDM overestimates ground-state correlations \cite{TTS} and shows instabilities of the obtained solutions \cite{toh12} 
for strongly interacting cases. Obviously the problems of TDDM originate in the
truncation scheme of the BBGKY hierarchy where the three-body density matrix is approximated by the one-body and two-body density matrices and true three-body
correlations are completely neglected. One way of overcoming the problems is to include the equation of motion for the three-body density matrix approximating
a four-body density matrix with lower-level density matrices. Such an attempt has been made for a model hamiltonian \cite{TS10} and showed that the results are drastically
improved by including genuine three-body correlations. However, it is impractical to treat explicitly the equation of motion for the three-body
density matrix in realistic cases. In this paper we show that an approximation for the three-body density matrix with the antisymmetrized products of the two-body density matrices
can simulate well the time-evolution of the three-body density matrix without explicitly solving its equation of motion. 
The paper is organized as follows; the formulation and the model hamiltonians are given in sect. 2, the results are presented in sect. 3
and sect. 4 is devoted to a summary.

\section{Formulation}
\subsection{Equations of motion for reduced density matrices}
We consider a system of $N$ fermions and assume that the hamiltonian $H$ consisting of a one-body part and a two-body interaction
\begin{eqnarray}
H=\sum_\alpha\epsilon_\alpha a^\dag_\alpha a_\alpha
+\frac{1}{2}\sum_{\alpha\beta\alpha'\beta'}\langle\alpha\beta|v|\alpha'\beta'\rangle
a^\dag_{\alpha}a^\dag_\beta a_{\beta'}a_{\alpha'},
\label{totalH}
\end{eqnarray}
where $a^\dag_\alpha$ and $a_\alpha$ are the creation and annihilation operators of a particle at
a single-particle state $\alpha$.

TDDM gives
the coupled equations of motion for the one-body density matrix (the occupation matrix) $n_{\alpha\alpha'}$
and the two-body density matrix $\rho_{\alpha\beta\alpha'\beta'}$.
These matrices are defined as
\begin{eqnarray}
n_{\alpha\alpha'}(t)&=&\langle\Phi(t)|a^\dag_{\alpha'} a_\alpha|\Phi(t)\rangle,
\\
\rho_{\alpha\beta\alpha'\beta'}(t)&=&\langle\Phi(t)|a^\dag_{\alpha'}a^\dag_{\beta'}
 a_{\beta}a_{\alpha}|\Phi(t)\rangle,
 \label{rho2}
\end{eqnarray}
where $|\Phi(t)\rangle$ is the time-dependent total wavefunction

\noindent
$|\Phi(t)\rangle=\exp[-iHt/\hbar] |\Phi(t=0)\rangle$.
The equations in TDDM are written as
\begin{eqnarray}
i \hbar\dot{n}_{\alpha\alpha'}&=&
(\epsilon_{\alpha}-\epsilon_{\alpha'}){n}_{\alpha\alpha'}
\nonumber \\
&+&\sum_{\lambda_1\lambda_2\lambda_3}
[\langle\alpha\lambda_1|v|\lambda_2\lambda_3\rangle \rho_{\lambda_2\lambda_3\alpha'\lambda_1}
\nonumber \\
&-&\rho_{\alpha\lambda_1\lambda_2\lambda_3}\langle\lambda_2\lambda_3|v|\alpha'\lambda_1\rangle],
\label{n2}
\end{eqnarray}
\begin{eqnarray}
i\hbar\dot{\rho}_{\alpha\beta\alpha'\beta'}&=&
(\epsilon_{\alpha}
+\epsilon_{\beta}
-\epsilon_{\alpha'}
-\epsilon_{\beta'}){\rho}_{\alpha\beta\alpha'\beta'}
\nonumber \\
&+&\sum_{\lambda_1\lambda_2}[
\langle\alpha\beta|v|\lambda_1\lambda_2\rangle\rho_{\lambda_1\lambda_2\alpha'\beta'}
\nonumber \\
&-&\langle\lambda_1\lambda_2|v|\alpha'\beta'\rangle\rho_{\alpha\beta\lambda_1\lambda_2}]
\nonumber \\
&+&\sum_{\lambda_1\lambda_2\lambda_3}
[\langle\alpha\lambda_1|v|\lambda_2\lambda_3\rangle\rho_{\lambda_2\lambda_3\beta\alpha'\lambda_1\beta'}
\nonumber \\
&+&\langle\lambda_1\beta|v|\lambda_2\lambda_3\rangle\rho_{\lambda_2\lambda_3\alpha\alpha'\lambda_1\beta'}
\nonumber \\
&-&\langle\lambda_1\lambda_2|v|\alpha'\lambda_3\rangle\rho_{\alpha\lambda_3\beta\lambda_1\lambda_2\beta'}
\nonumber \\
&-&\langle\lambda_1\lambda_2|v|\lambda_3\beta'\rangle\rho_{\alpha\lambda_3\beta\lambda_1\lambda_2\alpha'}],
\label{N3C2}
\end{eqnarray} 
where $\rho_{\alpha\beta\gamma\alpha'\beta'\gamma'}$ is a three-body density-matrix.
In TDDM the BBGKY hierarchy is truncated by approximating the three-body density matrix with
the antisymmetrized products of the one-body and two-body density matrices \cite{WC,GT}.
The TDDM equations eqs. (\ref{n2}) and (\ref{N3C2}) conserve the total number of particles $N=\sum_\alpha n_{\alpha\alpha}$
and the total energy \cite{WC,GT}
\begin{eqnarray}
E_{\rm tot}=\sum_\alpha\epsilon_\alpha n_{\alpha\alpha}+
\frac{1}{2}\sum_{\alpha\beta\alpha'\beta'}\langle\alpha\beta|v|\alpha'\beta'\rangle\rho_{\alpha'\beta'\alpha\beta}.
\end{eqnarray}

\subsection{Approximation for the three-body density matrix}
To separate the mean-field contributions,
it is convenient to introduce the two-body and three-body correlation matrices $C_{\alpha\beta\alpha'\beta'}$ 
and $C_{\alpha\beta\gamma\alpha'\beta'\gamma'}$ which describe genuine two-body and three-body correlations
such that
\begin{eqnarray}
\rho_{\alpha\beta\alpha'\beta'}&=&{\cal A}(n_{\alpha\alpha'}n_{\beta\beta'})+C_{\alpha\beta\alpha'\beta'},\\
\rho_{\alpha\beta\gamma\alpha'\beta'\gamma'}&=&{\cal A}(n_{\alpha\alpha'}n_{\beta\beta'}n_{\gamma\gamma'}
+n_{\alpha\alpha'}C_{\beta\gamma\beta'\gamma'})
\nonumber \\
&+&C_{\alpha\beta\gamma\alpha'\beta'\gamma'},
\end{eqnarray}
where ${\cal A}(\cdot \cdot)$ means that the quantities in the parenthesis are appropriately antisymmetrized.
In TDDM the three-body correlation matrix $C_{\alpha\beta\gamma\alpha'\beta'\gamma'}$ is neglected.

First we consider a perturbative expression for the three-body correlation matrix using
the following ground state $|Z\rangle$ \cite{jemai11} of coupled cluster theory
\begin{eqnarray}
|Z\rangle=e^Z|{\rm HF}\rangle
\end{eqnarray}
with 
\begin{eqnarray}
Z=\frac{1}{4}\sum_{pp'hh'}z_{pp'hh'}a^\dag_pa^\dag_{p'}a_{h'}a_h,
\end{eqnarray}
where $|{\rm HF}\rangle$ is the Hartree-Fock (HF) ground state and  
$z_{pp'hh'}$ is antisymmetric under the exchanges of $p \leftrightarrow p'$ and $h \leftrightarrow h'$.
Here, $p$ and $h$ refer to particle and hole states, respectively. 
In the lowest order of $z_{pp'hh'}$ the two-body correlation matrices are given by
\begin{eqnarray}
C_{pp'hh'}&\approx& z_{pp'hh'},\\
C_{hh'pp'}&\approx& z^*_{pp'hh'},
\end{eqnarray}
and the three-body correlation matrices by
\begin{eqnarray}
C_{p_1p_2h_1p_3p_4h_2}&\approx&\sum_{h}z^*_{p_3p_4hh_1}z_{p_1p_2h_2h},
\label{purt01}\\
C_{p_1h_1h_2p_2h_3h_4}&\approx&\sum_{p}z^*_{p_2ph_1h_2}z_{p_1ph_3h_4}.
\label{purt02}
\end{eqnarray}
These relations give the perturbative expressions for 

\noindent
$C_{\alpha\beta\gamma\alpha'\beta'\gamma'}$
in terms of $C_{pp'hh'}$
\begin{eqnarray}
C_{p_1p_2h_1p_3p_4h_2}&\approx&\sum_{h}C_{hh_1p_3p_4}C_{p_1p_2h_2h},
\label{purt1}\\
C_{p_1h_1h_2p_2h_3h_4}&\approx&\sum_{p}C_{h_1h_2p_2p}C_{p_1ph_3h_4}.
\label{purt2}
\end{eqnarray}
Similarly, in the second order of $z_{pp'hh'}$ the occupation matrix and other elements of the two-body correlation matrices are given by
\begin{eqnarray}
n_{hh}&\approx&1-\frac{1}{2}\sum_{pp'h'}C_{pp'hh'}C_{hh'pp'},
\label{hh1}\\
n_{pp}&\approx&\frac{1}{2}\sum_{hh'p'}C_{pp'hh'}C_{hh'pp'},
\label{pp1}\\
C_{p_1h_1p_2h_2}&\approx&\sum_{ph}C_{p_1ph_1h}C_{hh_1p_2p},
\label{ph2}\\
C_{p_1p_2p_3p_4}&\approx&\frac{1}{2}\sum_{hh'}C_{p_1p_2hh'}C_{hh'p_3p_4},
\label{pp2}\\
C_{h_1h_2h_3h_4}&\approx&\frac{1}{2}\sum_{pp'}C_{h_1h_2pp'}C_{pp'h_3h_4}.
\label{hh2}
\end{eqnarray}
Equations (\ref{purt1})-(\ref{purt2}) and (\ref{ph2})-(\ref{hh2}) imply that the three-body correlation matrix is of the same order of magnitude as some of
the two-body correlation matrix. From the above perturbative considerations we propose a truncation scheme of the BBGKY hierarchy
that in stead of neglecting the three-body correlation matrix we include them using eqs. (\ref{purt1}) and (\ref{purt2})
in the equation of motion for the two-body density matrix eq. (\ref{N3C2}).

There are identities which are satisfied by exact reduced density matrices:
\begin{eqnarray}
n_{\alpha\alpha'}&=&\frac{1}{N-1}\sum_\lambda\rho_{\alpha\lambda\alpha'\lambda}\\
\rho_{\alpha\beta\alpha'\beta'}&=&\frac{1}{N-2}\sum_\lambda\rho_{\alpha\beta\lambda\alpha'\beta'\lambda}\\
\rho_{\alpha\beta\gamma\alpha'\beta'\gamma'}&=&\frac{1}{N-3}\sum_\lambda\rho_{\alpha\beta\gamma\lambda\alpha'\beta'\gamma'\lambda}
\end{eqnarray}
and so on, where $\rho_{\alpha\beta\gamma\lambda\alpha'\beta'\gamma'\lambda}$ is a four-body density matrix.
Using the correlation matrices, the first three identities are explicitly written as \cite{TSW}
\begin{eqnarray}
n_{\alpha\alpha'}&=&\sum_{\lambda}(n_{\alpha\lambda}n_{\lambda\alpha'}-C_{\alpha\lambda\alpha'\lambda}),
\label{1body}\\
C_{\alpha\beta\alpha'\beta'}&=&\frac{1}{2}\sum_{\lambda}(n_{\alpha\lambda}C_{\lambda\beta\alpha'\beta'}
+n_{\beta\lambda}C_{\alpha\lambda\alpha'\beta'}
\nonumber \\
&+&n_{\lambda\alpha'}C_{\alpha\beta\lambda\beta'}
+n_{\lambda\beta'}C_{\alpha\beta\alpha'\lambda}
\nonumber \\
&-&C_{\alpha\beta\lambda\alpha'\beta'\lambda}),
\label{2body}\\
C_{\alpha\beta\gamma\alpha'\beta'\gamma'}&=&\frac{1}{3}\sum_\lambda(n_{\alpha\lambda}C_{\lambda\beta\gamma\alpha'\beta'\gamma'}
+n_{\beta\lambda}C_{\alpha\lambda\gamma\alpha'\beta'\gamma'}
\nonumber \\
&+&n_{\gamma\lambda}C_{\alpha\beta\lambda\alpha'\beta'\gamma'}
+n_{\lambda\alpha'}C_{\alpha\beta\gamma\lambda\beta'\gamma'}
\nonumber \\
&+&n_{\lambda\beta'}C_{\alpha\beta\gamma\alpha'\lambda\gamma'}
+n_{\lambda\gamma'}C_{\alpha\beta\gamma\alpha'\beta'\lambda}
\nonumber \\
&-&C_{\alpha\beta\alpha'\lambda}C_{\gamma\lambda\beta'\gamma'}
-C_{\alpha\beta\gamma'\lambda}C_{\gamma\lambda\alpha'\beta'}
\nonumber \\
&-&C_{\alpha\beta\lambda\beta'}C_{\gamma\lambda\alpha'\gamma'}
-C_{\alpha\gamma\alpha'\lambda}C_{\lambda\beta\beta'\gamma'}
\nonumber \\
&-&C_{\alpha\gamma\beta'\lambda}C_{\beta\lambda\alpha'\gamma'}
-C_{\alpha\gamma\lambda\gamma'}C_{\beta\lambda\alpha'\beta'}
\nonumber \\
&-&C_{\alpha\lambda\alpha'\beta'}C_{\beta\gamma\gamma'\lambda}
-C_{\alpha\lambda\alpha'\gamma'}C_{\beta\gamma\lambda\beta'}
\nonumber \\
&-&C_{\alpha\lambda\beta'\gamma'}C_{\beta\gamma\alpha'\lambda}-C_{\alpha\beta\gamma\lambda\alpha'\beta'\gamma'\lambda}),
\label{3body}
\end{eqnarray}
where $C_{\alpha\beta\gamma\lambda\alpha'\beta'\gamma'\lambda}$ is the four-body correlation matrix.
The TDDM equations without the three-body correlation matrix do not conserve the identity eq. (\ref{1body}) \cite{GT}.
 Suppose that the dominant two-body
correlation matrices are either two particle - two hole and two hole - two particle types and that the deviation 
of $n_{\alpha\alpha'}$ from the HF values are small. Then eq. (\ref{3body}) without $C_{\alpha\beta\gamma\lambda\alpha'\beta'\gamma'\lambda}$
is reduced to eqs. (\ref{purt1}) and (\ref{purt2}).
Similarly, it can be shown using eqs. (\ref{purt1}) and (\ref{purt2}) that the occupation matrices and 
the two-body correlation matrices given by eqs. (\ref{hh1})-(\ref{hh2}) satisfy eqs. (\ref{1body}) and (\ref{2body}).
We also test the following truncation scheme:
Neglecting the four-body correlation matrix $C_{\alpha\beta\gamma\delta\alpha'\beta'\gamma'\delta'}$ in eq. (\ref{3body}), 
we use the three-body correlation matrix
given by eq. (\ref{3body}) in the equation of motion for the two-body density matrix.
For simplicity we do not consider the off-diagonal elements of $n_{\alpha\alpha'}$ in the applications shown below.
We show that eqs. (\ref{purt1}) and (\ref{purt2}) give better results than eq. (\ref{3body}) in the case of the Lipkin model, 
indicating that
the use of the fractional occupation in eq. (\ref{3body}) sometimes overestimates the effects of the three-body correlation matrix
when the four-body correlation matrix is neglected.

The ground state in TDDM is given as a stationary solution of the TDDM equations 
(eqs. (\ref{n2}) and (\ref{N3C2})). 
We use the following adiabatic method to obtain a nearly stationary
solution \cite{adiabatic1}: Starting from a non-interacting configuration,
we solve eqs. (\ref{n2}) and (\ref{N3C2}) gradually increasing the interaction 
$v({\bm r})\times t/T$. To suppress oscillating components which come from the mixing
of excited states, we must take large $T$.
This method is based on the Gell-Mann-Low adiabatic theorem \cite{gell} and 
has often been used to obtain correlated ground states \cite{pfitz,lacroix}.
 
\subsection{Model hamiltonians}
We apply our truncation schemes to the following three cases where comparison with the exact solutions can be made.
\subsubsection{Lipkin mdel}
The Lipkin model \cite{Lip} describes an $N$-fermions system with two
$N$-fold degenerate levels with energies $\epsilon/2$ and $-\epsilon/2$,
respectively. The upper and lower levels are labeled by quantum number
$p$ and $-p$, respectively, with $p=1,2,...,N$. We consider
the standard hamiltonian
\begin{equation}
\hat{H}=\epsilon \hat{J}_{z}+\frac{V}{2}(\hat{J}_+^2+\hat{J}_-^2),
\label{elipkin}
\end{equation}
where the operators are given as
\begin{eqnarray}
\hat{J}_z&=&\frac{1}{2}\sum_{p=1}^N(a_p^{\dag}a_p-{a_{-p}}^{\dag}a_{-p}), \\
\hat{J}_{+}&=&\hat{J}_{-}^{\dag}=\sum_{p=1}^N a_p^{\dag}a_{-p}.
\end{eqnarray}

\subsubsection{Hubbard model}
To test our truncation schemes for a case which involves more single-particle states than the Likin model,
we consider the one-dimensional (1D) Hubbard model with periodic boundary conditions.
In momentum space the hamiltonian is given by
\begin{eqnarray}
H&=&\sum_{{\bm k},\sigma}\epsilon_{k}a^\dag_{{\bm k},\sigma}a_{{\bm k},\sigma}
\nonumber \\
&+&\frac{U}{2N}\sum_{{\bm k},{\bm p},{\bm q},\sigma}
a^\dag_{{\bm k},\sigma}a_{{\bm k}+{\bm q},\sigma}a^\dag_{{\bm p},-\sigma}a_{{\bm p}-{\bm q},-\sigma},
\end{eqnarray}
where 
$U$ is the on-site Coulomb matrix element, $\sigma$ spin projection and 
the single-particle energies are given by $\epsilon_k=-2t\sum_{d=1}^D\cos(k_d)$
with the nearest-neighbor hopping potential $t$.
We consider the case of the six sites at half filling.
In the first Brillouin zone $-\pi\le k <\pi$ there are the following wave numbers
\begin{eqnarray}
k_1&=&0,~~~k_2=\frac{\pi}{3},~~~k_3=-\frac{\pi}{3},
\nonumber \\
k_4&=&\frac{2\pi}{3},~~~k_5=-\frac{2\pi}{3}.~~~k_6=-\pi.
\end{eqnarray}
The single-particle energies are $\epsilon_1=-2t$, $\epsilon_2=\epsilon_3=-t$, 
$\epsilon_4=\epsilon_5=t$ and $\epsilon_6=2t$.

\subsubsection{Dipolar fermion gas} 
As a more realistic case,
we consider a magnetic dipolar gas of fermions with spin one half, 
which is trapped in a spherically symmetric harmonic potential with
frequency $\omega$. The system is described by the hamiltonian 
\begin{eqnarray}
H=\sum_\alpha\epsilon_\alpha a^\dag_\alpha a_\alpha
+\frac{1}{2}\sum_{\alpha\beta\alpha'\beta'}\langle\alpha\beta|v|\alpha'\beta'\rangle
a^\dag_{\alpha}a^\dag_\beta a_{\beta'}a_{\alpha'},
\label{totalH1}
\end{eqnarray}
where $a^\dag_\alpha$ and $a_\alpha$ are the creation and annihilation operators of an atom at
a harmonic oscillator state $\alpha$ 
corresponding to the trapping potential $V(r)=m\omega^2r^2/2$ and
$\epsilon_\alpha=\omega(n+3/2)$ with $n=0,~1,~2,....$.
We assume that $\alpha$ contains the spin quantum number $\sigma$.
In eq. (\ref{totalH1}) $\langle\alpha\beta|v|\alpha'\beta'\rangle$ is the matrix element of 
a pure magnetic dipole-dipole interaction \cite{dipole}
\begin{eqnarray}
v(r)&=&-\frac{1}{r^3}\left(3({\bm d}_1\cdot\hat{\bm r})({\bm d}_2\cdot\hat{\bm r})
-{\bm d}_1\cdot{\bm d}_2\right)
\nonumber \\
&-&\frac{8\pi}{3}{\bm d}_1\cdot{\bm d}_2\delta^3({\bm r}),
\nonumber \\
\label{vdd}
\end{eqnarray}
where ${\bm d}$ is the magnetic dipole moment, ${\bm r}={\bm r}_1-{\bm r}_2$ and $\hat{\bm r}={\bm r}/r$.
The magnetic dipole moment for spin 1/2 is given by ${\bm d}=d{\bm \sigma}$ where ${\bm \sigma}$ is Pauli matrix.
In the case of completely polarized gases 
the second term on the right-hand side of eq. (\ref{vdd}) can be neglected because the exchange term
cancels out the direct term. The contact term (the second term on the right-hand side of eq. (\ref{vdd})) is usually
omitted in the study of dipolar gases. However, it is well-known that the contact term for the proton and electron magnetic 
dipole moments is essential to explain the hyperfine splitting of
a hydrogen atom. Therefore, in the following calculations we keep it as it is.
We consider the $N=8$ case where noninteracting HF ground state consists of 
the fully occupied $1s$ and $1p$ harmonic oscillator states.  
For simplicity we take only the $2s$ and $1d$ states as particle states and keep the $1s$ state frozen,
i.e., only the $1p$ state is active as a hole-state.

\begin{figure}
\resizebox{0.5\textwidth}{!}{%
 \includegraphics{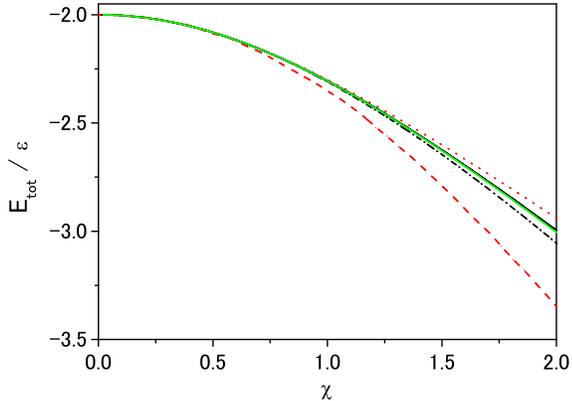}
}

\caption{Ground-state energies in TDDM with the three-body correlation matrix given by eq. (\ref{3body}) (dotted line) and by eqs. (\ref{purt1}) and (\ref{purt2})
(green (gray) solid line)
as a function of $\chi=(N-1)|V|/\epsilon$ for $N=4$. 
The dashed line depicts the results in TDDM where the three-body correlation matrix is neglected, 
and the dot-dashed line the exact values. 
The results with the three-body correlation matrix obtained from solving the equation of motion are given by the black
solid line.}
\label{lip1}
\end{figure}

\begin{figure}
\resizebox{0.5\textwidth}{!}{%
\includegraphics{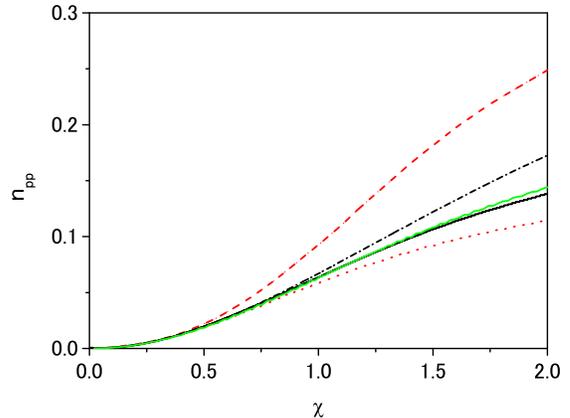}
}
\caption{Occupation probability of the upper state as a function of $\chi$ for $N=4$. The meaning of the five
lines is the same as in fig. \ref{lip1}.}
\label{lip2}
\end{figure}

\begin{figure}
\resizebox{0.5\textwidth}{!}{%
\includegraphics{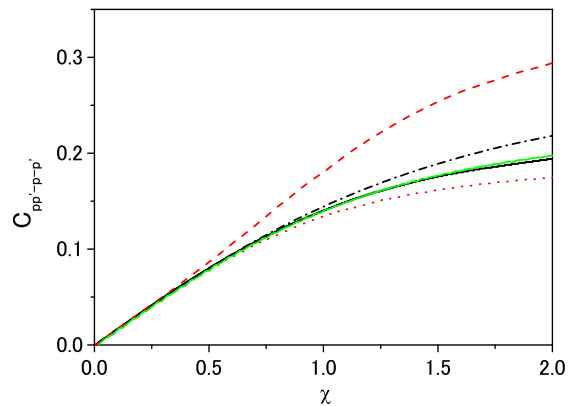}
}
\caption{Two-body correlation matrix $C_{pp'-p-p'}$ as a function of $\chi$ for $N=4$.
The meaning of the five
lines is the same as in fig. \ref{lip1}.}
\label{lip3}
\end{figure}

\section{Results}
\subsection{Lipkin model}
The ground-state energies obtained from various approximations for the three-body correlation matrix are shown
in fig. \ref{lip1} as a function of $\chi=(N-1)|V|/\epsilon$ for $N=4$. 
The RPA solution becomes unstable at $\chi=1$.
The dashed line shows the results in TDDM where the three-body correlation matrix is neglected.
The dotted and green (gray) solid lines are for the results with the
three-body correlation matrix given by eq. (\ref{3body}) and by eqs. (\ref{purt1}) and (\ref{purt2}), respectively.
The results with the three-body correlation matrix obtained from solving the equation of motion are given by the black
solid line \cite{TS10}. The exact solutions are given by the dot-dashed line.
The occupation probability of the upper state and the two-body correlation matrix
$C_{pp'-p-p'}$ are shown in figs. \ref{lip2} and \ref{lip3}, respectively.
It is clear from figs. \ref{lip1}-\ref{lip3} that the neglect of the three-body correlation matrix
overestimates the ground-state correlations. The perturbative treatment eqs. (\ref{purt1}) and (\ref{purt2}) 
seems to give better results than the approximation eq. (\ref{3body}) which overly suppresses ground-state correlations.
Comparing with the results obtained with the HF values of $n_{\alpha\alpha}$, 
we found that the use of the fractional occupation $n_{\alpha\alpha}$ in eq. (\ref{3body}) causes this over suppression.
We also found that the omission of $C_{php'h'}$ in eq. (\ref{N3C2}) does not bring serious overestimation of the ground-state correlations even if
the three-body correlation matrix is excluded. This means that the three-body correlation matrix plays an important 
role in suppressing the particle - hole correlations. This is true for the other two cases of the model hamiltonians.
\begin{figure}
\resizebox{0.5\textwidth}{!}{%
\includegraphics{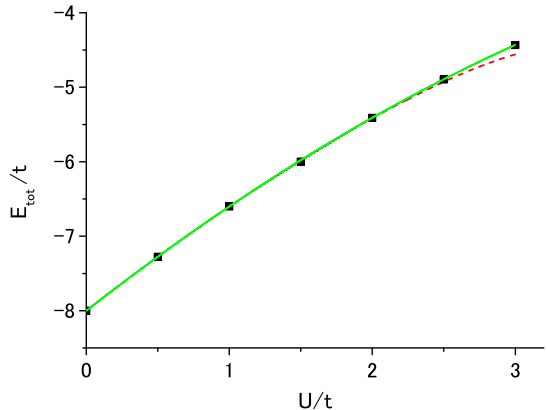}
}
\caption{Ground-state energy in TDDM (solid line) with $C_{\alpha\beta\gamma\alpha'\beta'\gamma'}$ given by eqs. (\ref{purt1}) and (\ref{purt2}) 
as a function of $U/t$ for the half-filling six-site Hubbard model.
The dashed line depicts the results in TDDM where the three-body correlation matrix is neglected, 
and the squares the exact values.}
\label{hub1}
\end{figure}

\subsection{Hubbard model}
\begin{figure}
\resizebox{0.5\textwidth}{!}{%
\includegraphics{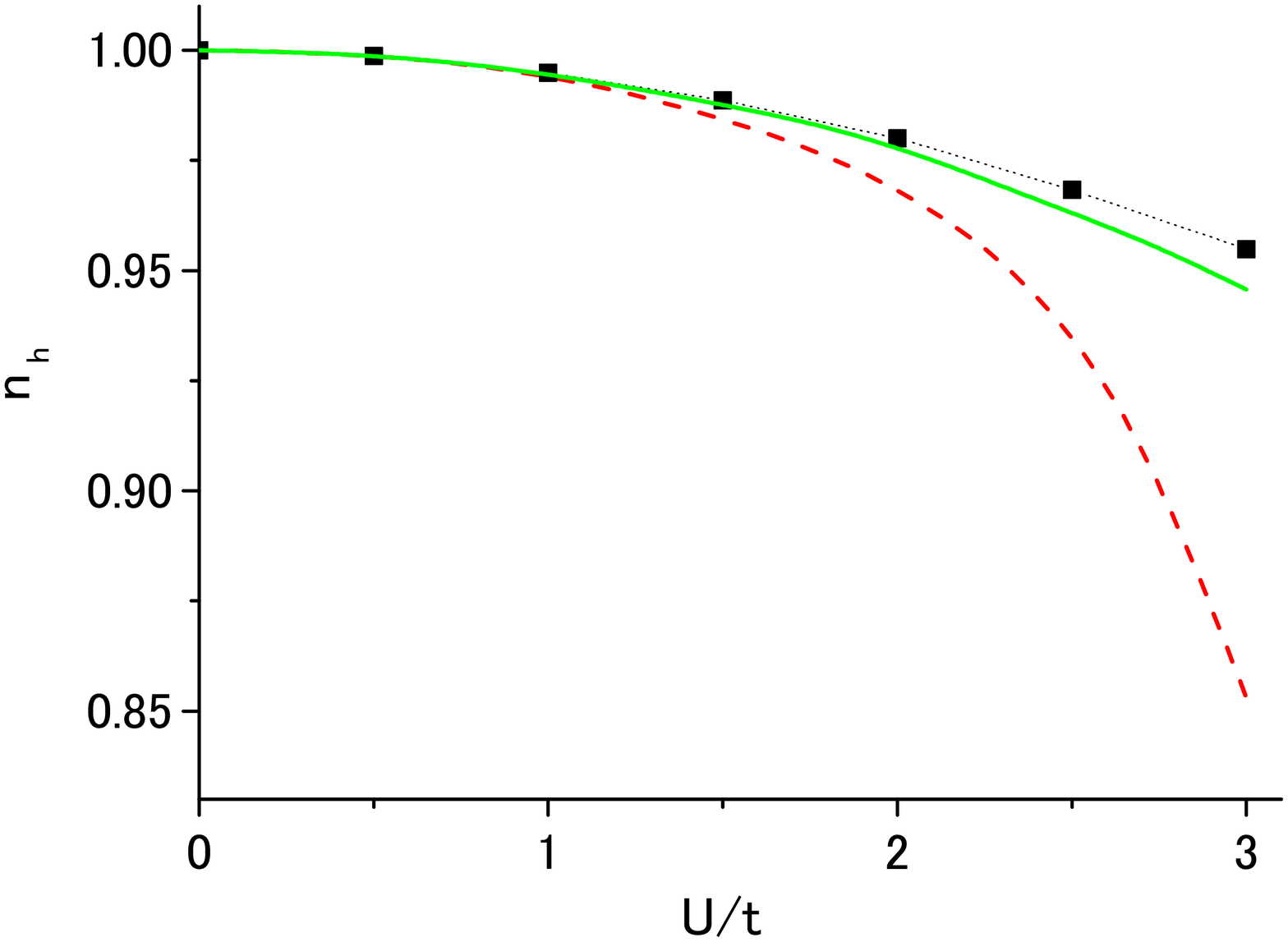}
}
\caption{Occupation probability of the initially occupied state with $k_1$ as a function of $U/t$. The meaning of the lines and symbol
is the same as in fig. \ref{hub1}.}
\label{hub21}
\end{figure}

\begin{figure}
\resizebox{0.5\textwidth}{!}{%
\includegraphics{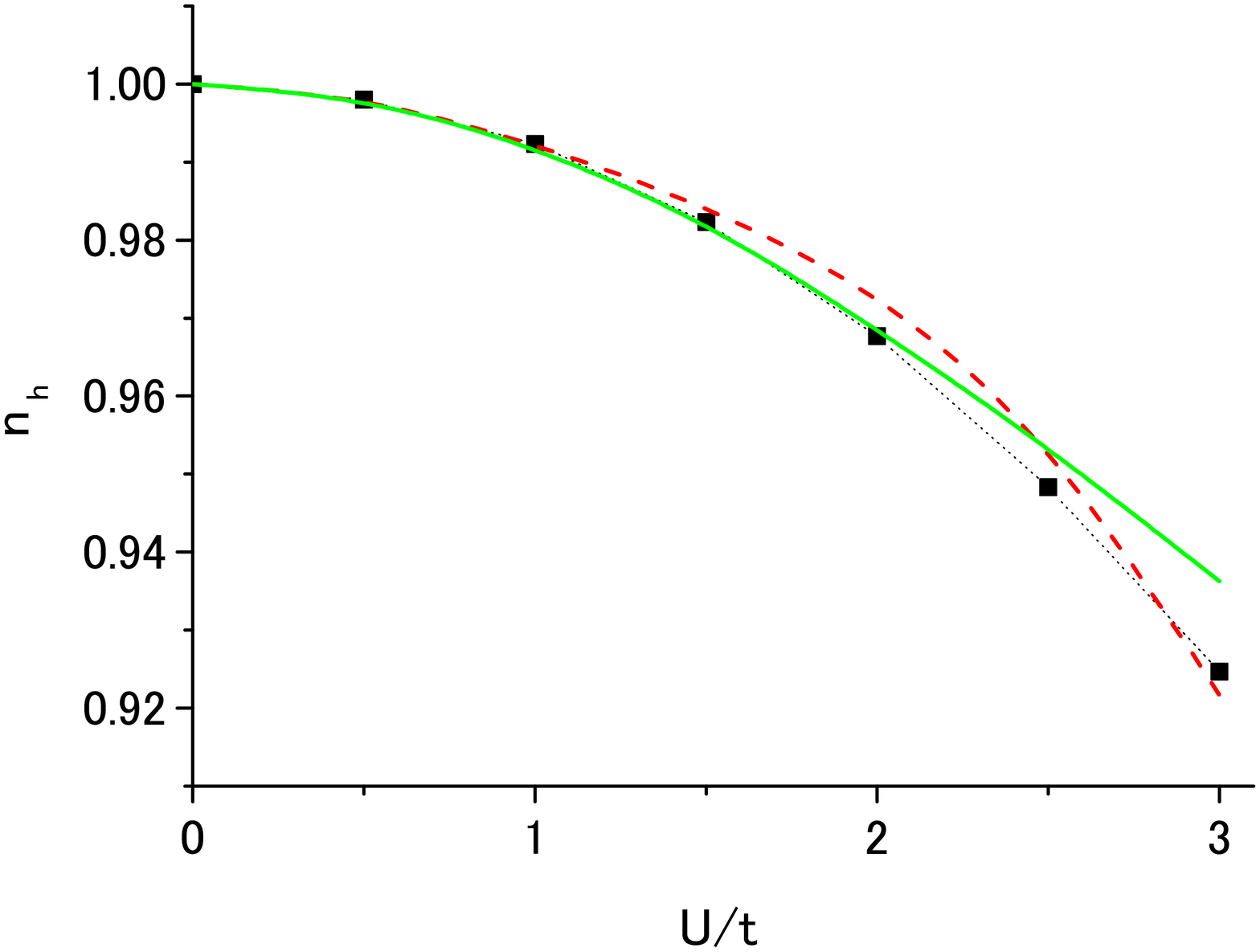}
}
\caption{Occupation probability of the second occupied state with $k_2$ as a function of $U/t$. The meaning of the lines and symbol
is the same as in fig. \ref{hub1}.}
\label{hub2}
\end{figure}

\begin{figure}
\resizebox{0.5\textwidth}{!}{
\includegraphics{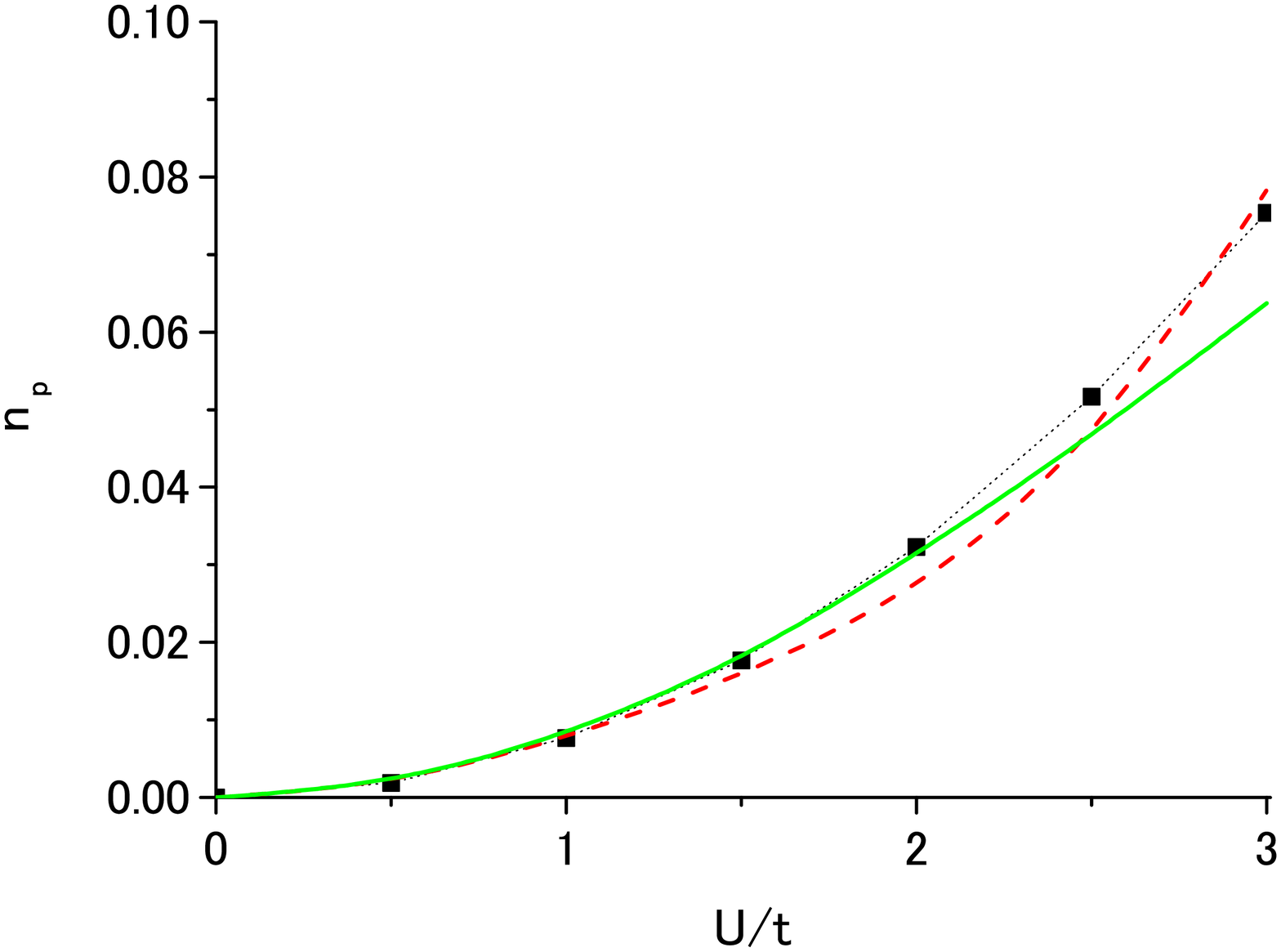}
}
\caption{Occupation probability of the first unoccupied state with $k_4$ as a function of $U/t$. The meaning of the lines and symbol
is the same as in fig. \ref{hub1}.}
\label{hub3}
\end{figure}

\begin{figure}
\resizebox{0.5\textwidth}{!}{%
\includegraphics{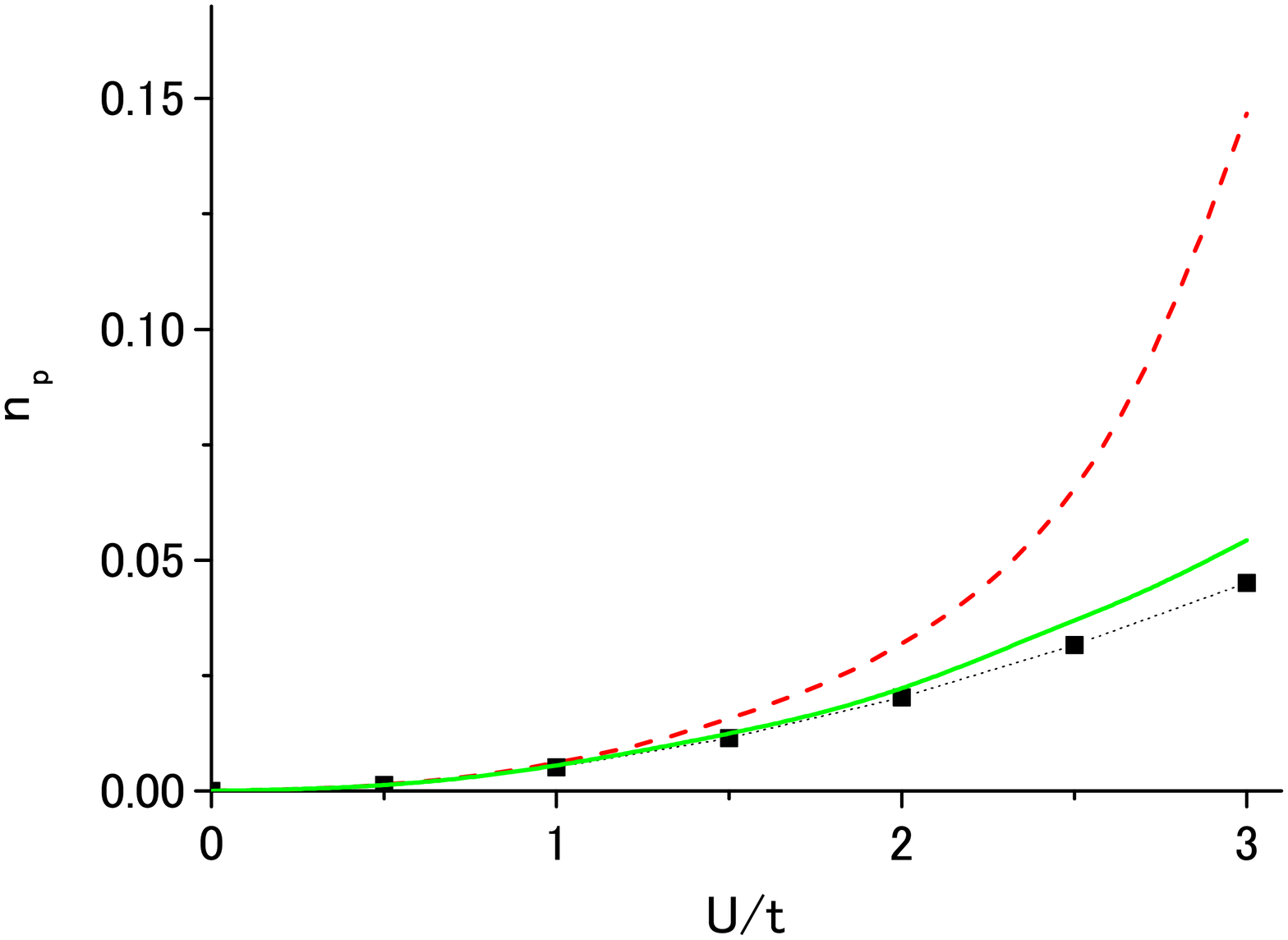}
}
\caption{Occupation probability of the last unoccupied state with $k_6$ as a function of $U/t$. The meaning of the lines and symbol
is the same as in fig. \ref{hub1}.}
\label{hub31}
\end{figure}

\begin{figure}
\resizebox{0.5\textwidth}{!}{%
\includegraphics{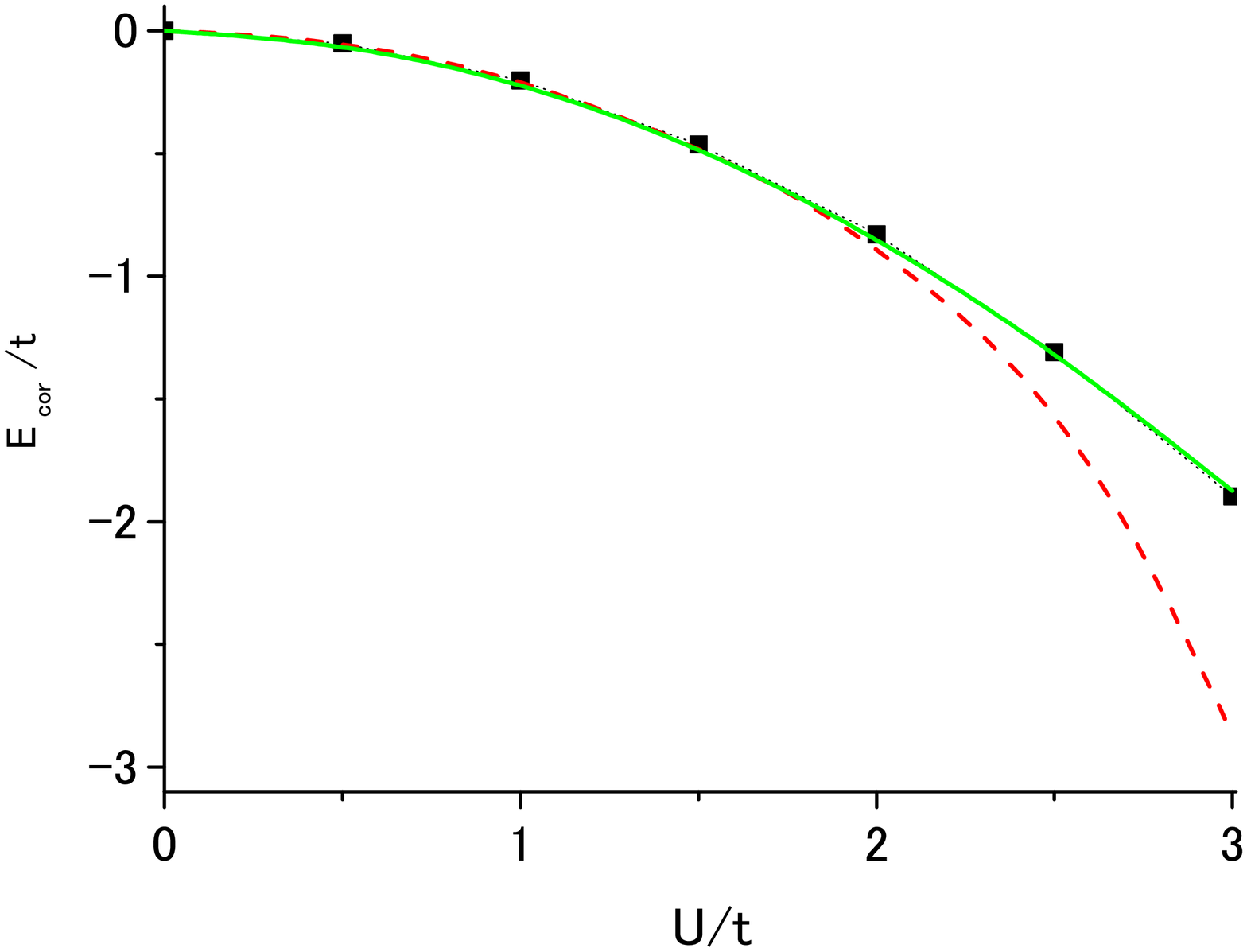}
}
\caption{Correlation energy $E_{\rm cor}$ as a function of $U/t$. The meaning of the lines and symbol
is the same as in fig. \ref{hub1}.}
\label{hubcor}
\end{figure}

The total energy calculated in TDDM with the three-body correlation matrix given by eqs. (\ref{purt1}) and (\ref{purt2}) (solid line) is shown 
in fig. \ref{hub1} as a function of $U/t$ for the six-site Hubbard model with half-filling.
The results obtained with eq. (\ref{3body}) are similar and not shown.
The RPA solution becomes unstable at $U/t=2.4$ \cite{jemai05}.
The dashed line shows the results in TDDM without the three-body correlation matrix. The exact solutions are given by the
squares. The occupation probabilities of the single-particle states are shown in figs. \ref{hub21} - \ref{hub31}.
Since it is not convenient to show each matrix element of $C_{\alpha\beta\alpha'\beta'}$, we show
the correlation energy given by $E_{\rm cor}=\sum_{\alpha\beta\alpha'\beta'}\langle\alpha\beta|v|\alpha'\beta'\rangle C_{\alpha'\beta'\alpha\beta}/2$
in fig. \ref{hubcor}.
Figures \ref{hub1}-\ref{hub21} and \ref{hub31}-\ref{hubcor} show that the neglect of the three-body correlation matrix overestimate the ground-state correlations
and that the approximation given by eqs. (\ref{purt1}) and (\ref{purt2}) improves the results.
The differences among the three calculations are not so large in the ground-state energy as in the correlation energy.
This is due to the fact that the increase in the correlation energy is compensated by that in the mean-field energy.

\begin{figure}
\resizebox{0.5\textwidth}{!}{%
\includegraphics{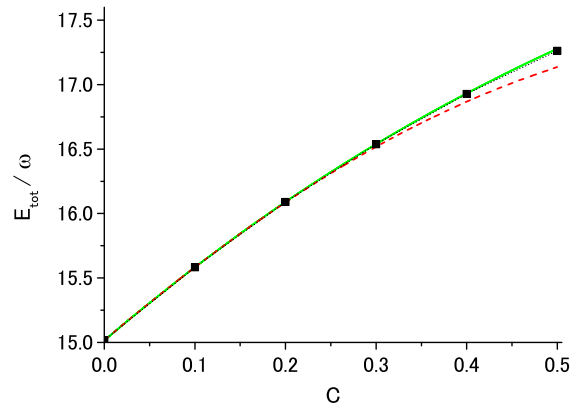}
}
\caption{Ground-state energy in TDDM (solid line) with $C_{\alpha\beta\gamma\alpha'\beta'\gamma'}$ given by eqs. (\ref{purt1}) and  (\ref{purt2})
as a function of $C$ for the $N=8$ dipolar fermion gas.
The dashed line depicts the results in TDDM where the three-body correlation matrix is neglected, 
and the squares the exact values. }
\label{dip1}
\end{figure}
\begin{figure}
\resizebox{0.5\textwidth}{!}{%
\includegraphics{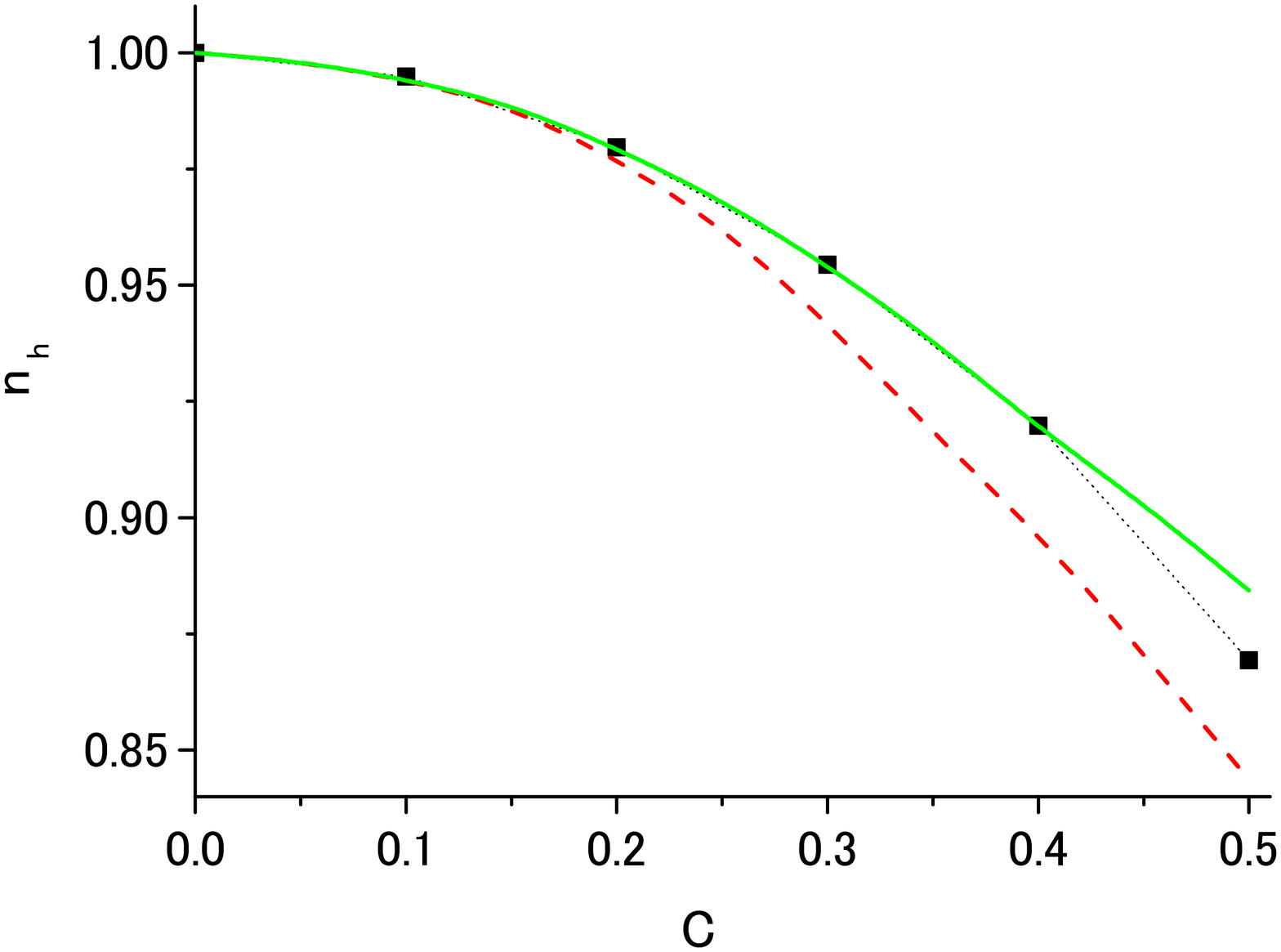}
}
\caption{Occupation probability of the $1p$ state with $m=1$ and $\sigma=-1/2$ as a function of $C$. The meaning of the lines and symbol
is the same as in fig. \ref{dip1}.}
\label{dip2}
\end{figure}
\begin{figure}
\resizebox{0.5\textwidth}{!}{%
\includegraphics{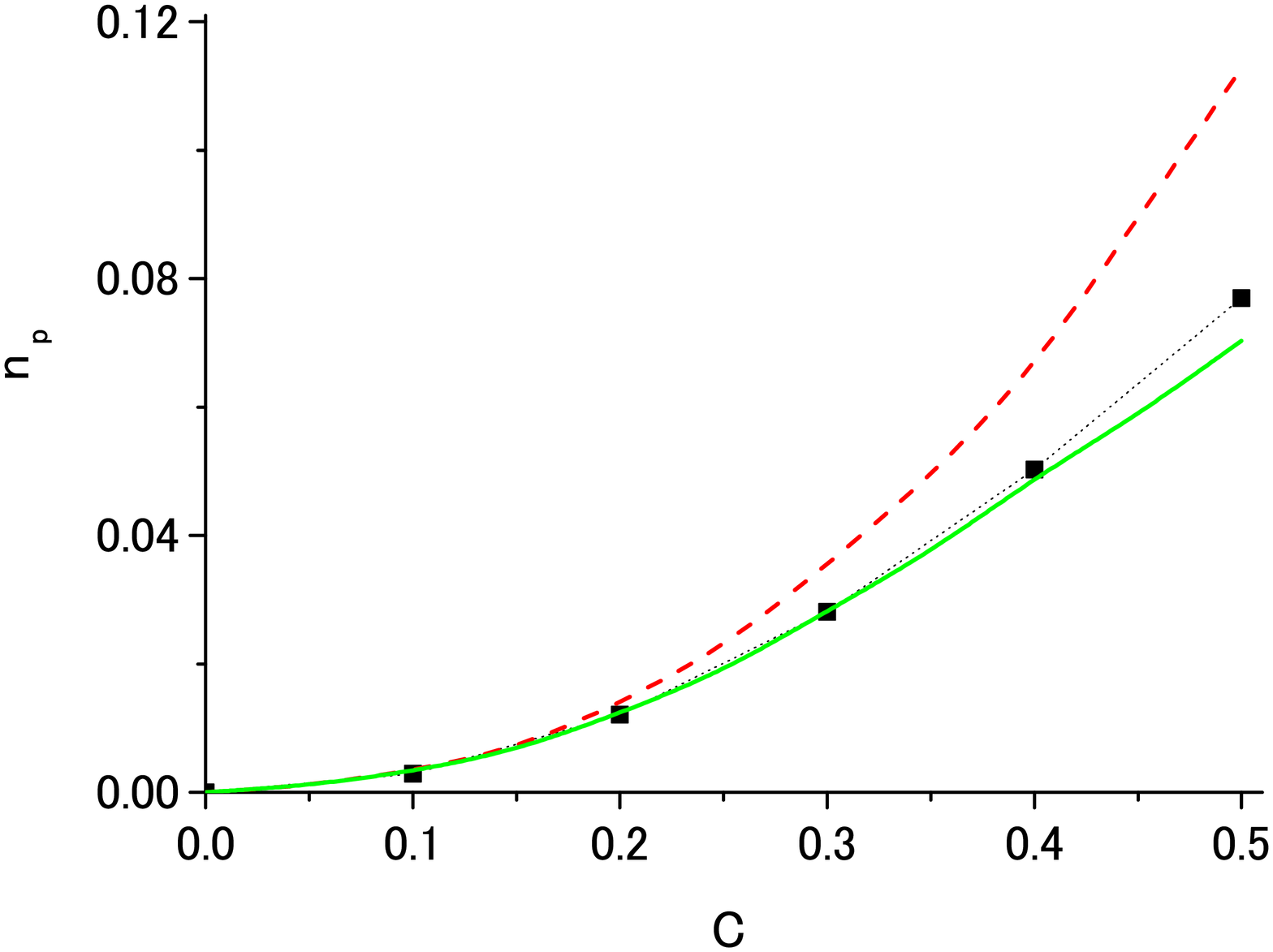}
}
\caption{Occupation probability of the $1d$ state with $m=2$ and $\sigma=-1/2$ as a function of $C$. The meaning of the lines and symbol
is the same as in fig. \ref{dip1}.}
\label{dip3}
\end{figure}
\begin{figure}
\resizebox{0.5\textwidth}{!}{%
\includegraphics{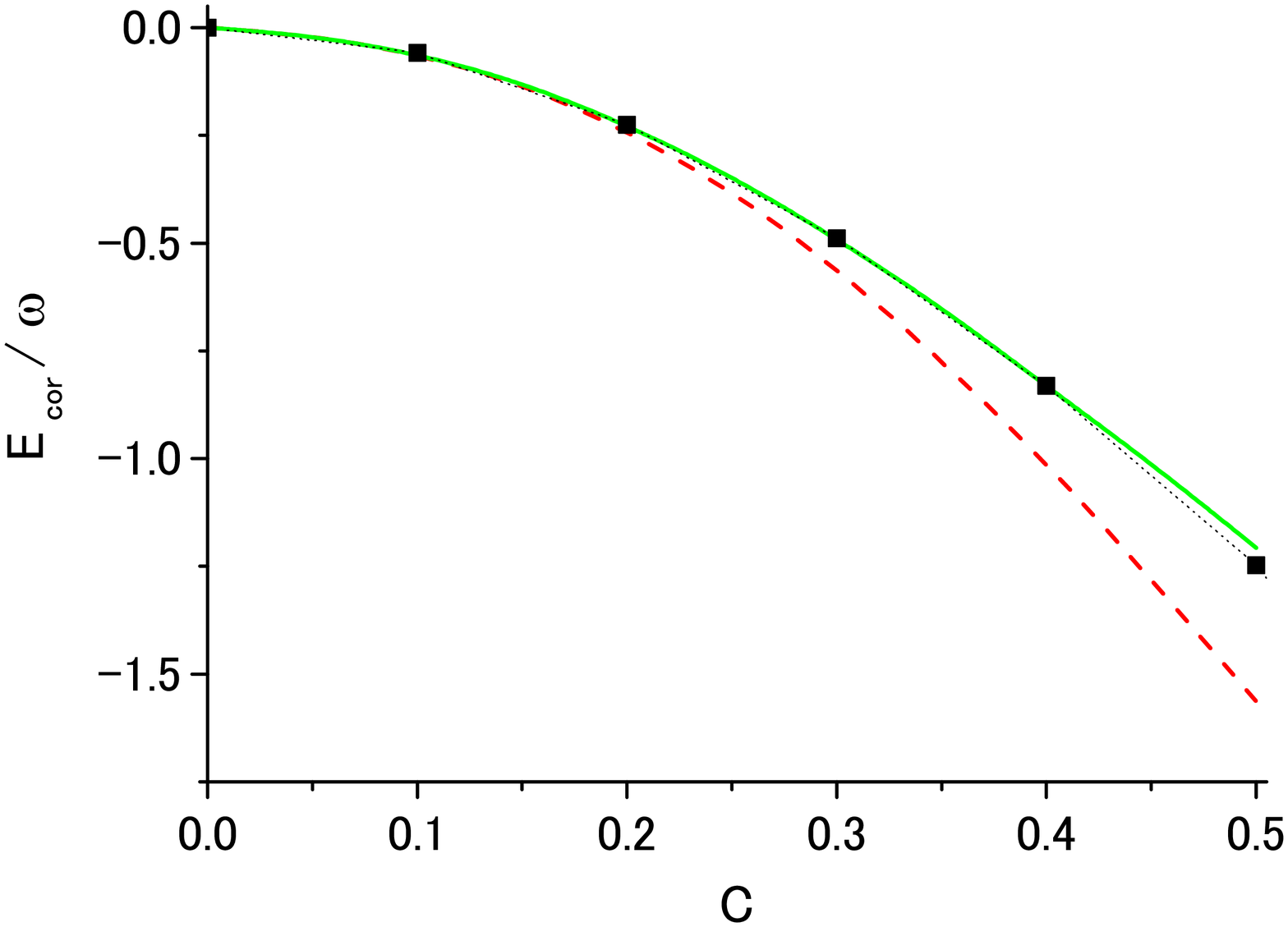}
}
\caption{Correlation energy $E_{\rm cor}$ as a function of $C$. The meaning of the lines and symbol
is the same as in fig. \ref{dip1}.}
\label{dipcor}
\end{figure}
\begin{figure}
\resizebox{0.5\textwidth}{!}{%
\includegraphics{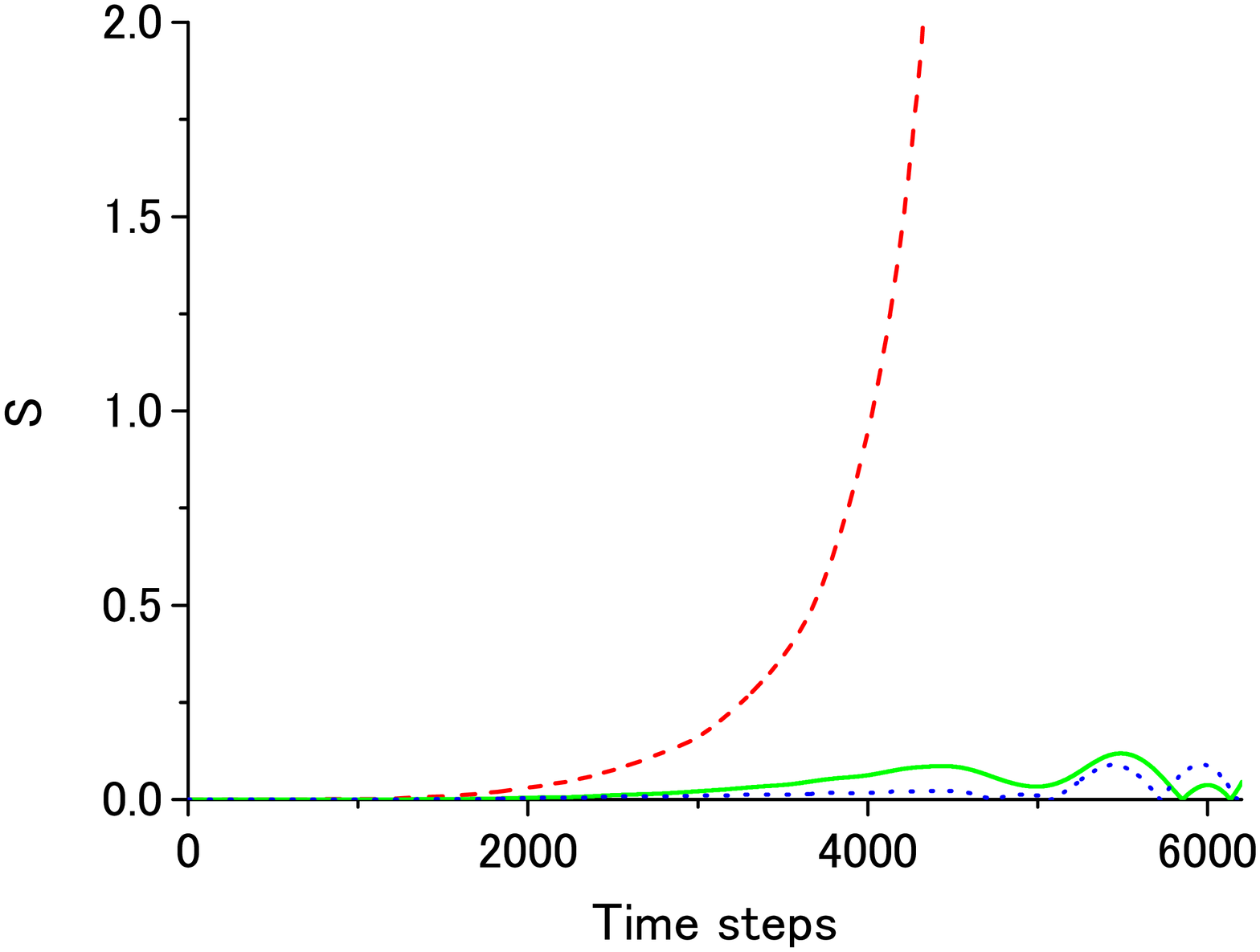}
}
\caption{Sum of the absolute value of eq.(\ref{1body}) as a function of time steps at $U/t=3$ for the six-site Hubbard model.
The dotted line and the green line show the results in TDDM with the three-body correlation matrix given by eq. (\ref{3body}) and by
eqs. (\ref{purt1}) and (\ref{purt2}) while the results without the three-body correlation matrix are given by the dashed line.}
\label{hub4}
\end{figure}

\subsection{Dipolar atomic gas} 
The total energy calculated in TDDM with the three-body correlation matrix given by eqs. (\ref{purt1}) and  (\ref{purt2}) is shown by the solid line
in fig. \ref{dip1} as a function of $C=d^2/\hbar\omega\xi^3$ for the $N=8$ dipolar fermion gas. Here $\xi$ is the oscillator 
length given by $\xi=\sqrt{\hbar/m\omega}$. 
The dashed line shows the results in TDDM without the three-body correlation matrix. The exact solutions are given by the
squares. The RPA solution becomes unstable for a spin-orbit coupling mode at $C=0.34$ \cite{toh13}.
Due to the spin-orbit coupling in the tensor force, the magnetic quantum number $m$ of orbital angular momentum and $\sigma$ 
are not good quantum numbers, and $n_{\alpha\alpha'}$ has nonvanishing off-diagonal elements.
In the application of eq. (\ref{3body}) we neglected the off-diagonal elements because they are small.
The results with eq. (\ref{3body}) are similar to those with eqs. (\ref{purt1}) and  (\ref{purt2}) and not shown here.
The occupation probabilities of the $1p$ state with $m=1$ and $\sigma=-1/2$ and the $1d$ state 
with $m=2$ and $\sigma=-1/2$ are shown in figs. \ref{dip2} and \ref{dip3}, respectively.
The correlation energy $E_{\rm cor}$ is shown in fig. \ref{dipcor}.
Figures \ref{dip1}-\ref{dipcor} again show that the neglect of the three-body correlation matrix overestimates the ground-state correlations
and that the approximation  eqs. (\ref{purt1}) and  (\ref{purt2}) drastically improves the results of the occupation probabilities and
the correlation energy.

\subsection{Stability of the ground state solution}
In the above we demonstrated that 
the inclusion of the three-body correlation matrix improves the ground-state properties in TDDM. We now show 
that it also stabilizes long-time behavior of the ground state.
To show the stability of the ground-state solution in TDDM, we present the sum $S$ of the absolute value of eq. (\ref{1body}) 
$S=\sum_{\alpha\alpha'}|n_{\alpha\alpha'}-\sum_{\lambda}(n_{\alpha\lambda}n_{\lambda\alpha'}-C_{\alpha\lambda\alpha'\lambda})|$
in fig. \ref{hub4} for the Hubbard model. In this model $n_{\alpha\alpha'}$ has no off-diagonal elements.
The time step 4000 corresponds to $T$ and for $t>T$ the interaction strength is fixed at $U/t=3$.
The dotted and solid lines show the results in TDDM with the three-body correlation matrix given by eq. (\ref{3body}) and by
eqs. (\ref{purt1}) and (\ref{purt2}) while the results without the three-body correlation matrix are given by the dashed line.
Since the ground-state solution obtained in the adiabatic approach inevitably contains small time-dependent components,
it has time evolution. If
the gradient method ref. \cite{toh07} is used to obtain a true stationary solution of the TDDM equations, the occupation matrix and the two-body
correlation matrix have no time evolution and $S$ stays constant though the identity eq. (\ref{1body}) is violated.
The violation of the identity eq. (\ref{1body}) 
causes such a serious problem for strong interactions that $n_{\alpha\alpha}$ exceeds unity or becomes negative.
When $n_{\alpha\alpha}$ become unphysical, the TDDM solution is no longer stable as shown in fig. \ref{hub4}. The three-body correlation
matrix plays a role in approximately conserving the identity eq. (\ref{1body}) and stabilizing the time evolution.
We found that neglect of the two-body correlation matrices of three particle - one hole and three hole - one particle types
such as $C_{pp'p''h}$ and $C_{hh'h''p}$ also stabilizes the time evolution in the Hubbard model even if the three-body correlation matrix is neglected.

\section{Summary}
We proposed truncation schemes of the time-dependent density-matrix approach in which the three-body density matrix is approximated
in terms of the squares of the two-body density matrices. We applied them to the ground states of the Lipkin model, the Hubbard model
and the trapped dipolar fermion gas and compared with the exact solutions. It was shown that the truncation schemes give better 
results than the approach without the
three-body density matrix. It was pointed out that the three-body correlation matrix plays a role in suppressing the particle - hole correlations.
It was also shown that such approximations for the three-body density matrix also drastically improve the
stability of the ground-state solutions because of approximate conservation of the identity for the 
one-body and two-body density matrices. Thus it was found that our density-matrix approach supplemented by the truncation schemes for the three-body density matrix
gives good approximation for the total ground-state wavefunction.
TDDM is now on good grounds for realistic approximations.


\begin{thebibliography}{99}
\bibitem{WC}
S. J. Wang and W. Cassing, Ann. Phys. {\bf 159}, 328 (1985).
\bibitem{GT}
M. Gong and M. Tohyama, Z. Phys. A{\bf 335}, 153 (1990).
\bibitem{TTS}
S. Takahara, M. Tohyama and P. Schuck, Phys. Rev. C{\bf 70}, 057307 (2004).
\bibitem{toh07}
M. Tohyama, Phys. Rev. C {\bf 75}, 044310 (2007).
\bibitem{pfitz}
A. Pfitzner, W. Cassing, and A. Peter, Nucl. Phys. A{\bf 577}, 753 (1994).
\bibitem{lacroix}
M. Assi$\acute{\rm e}$ and D. Lacroix, Phys. Rev. Lett. {\bf 102}, 202501 (2009).
\bibitem{toh12}
M. Tohyama, J. Phys. Soc. Jpn. {\bf 81}, 054707 (2012).
\bibitem{TS10}
M. Tohyama and P. Schuck, Eur. Phys. J. A {\bf 45}, 257 (2010).
\bibitem{jemai11}
M. Jema$\ddot{\rm i}$ and P. Schuck,
Atomic Nuclei {\bf 74}, N0. 8, 1139 (2011).
\bibitem{TSW}
M. Tohyama, P. Schuck and S. J. Wang, Z. Phys. A {\bf 339}, 341 (1991).
\bibitem{adiabatic1}
M. Tohyama, Phys. Rev. A{\bf 71}  (2005) 043613. 
\bibitem{gell}
M. Gell-Mann and F. Low, Phys. Rev. {\bf 84}, 350 (1951).
\bibitem{Lip} 
H. J. Lipkin, N. Meshkov and A. J. Glick, Nucl. Phys. {\bf 62}, 188 (1965).
\bibitem{dipole}
V. D. Barger and M. G. Olsson, {\it Classical electricity and magnetism} (Allyn and Bacon, Boston, 1987).
\bibitem{jemai05}
M. Jema$\ddot{{\rm i}}$, P. Schuck, J. Dekelsky and R. Bennaceur, Phys. Rev. B {\bf 71}, 085115 (2005).
\bibitem{toh13}
M. Tohyama, J. Phys. Soc. Jpn. {\bf 82}, 124004 (2013).
\end{thebibliography}
\end{document}